\documentclass[12pt]{article}
\usepackage[utf8]{inputenc}
\usepackage[T1]{fontenc}
\usepackage{microtype}
\usepackage[dvips]{graphicx}
\usepackage{xcolor}
\usepackage{times}
\usepackage{amsmath,amsfonts,amssymb,amsthm,epsfig,epstopdf,titling,url,array}
\usepackage{float}
\usepackage{subfig}
\usepackage{verbatim}

\usepackage{soul}
\usepackage[
breaklinks=true,colorlinks=true,	
linkcolor=black,urlcolor=black,citecolor=black,
bookmarks=true,bookmarksopenlevel=2]{hyperref}

\newcounter{lastnote}

\title{Evolution and metric signature change of maximally symmetric spaces under the Ricci flow}

\author
{R. Cartas-Fuentevilla, A. Herrera-Aguilar,  J. A. Olvera-Santamar\'ia\\\\
\normalsize{Instituto de F\'isica-LRT, Benem\'erita Universidad Aut\'onoma de Puebla,72570 Puebla, Pue.}\\
\normalsize{ rcartas@ifuap.buap.mx, aherrera@ifuap.buap.mx, jose.olverasantamaria@monash.edu}
}  
\date{}
\begin{document} 
\maketitle

\begin{abstract}
In this work we present solutions to the Ricci flow equations in arbitrary dimensions, particularizing for the $3d$ and $4d$ cases. We start by considering the $3d$ case and 
note that our solutions belong to the family of maximally symmetric spaces that can be extended to the $d\geq 4$ case following an analogue treatment. These solutions can be divided into two scenarios: maximally symmetric spaces with positive curvature i.e. de Sitter spaces, and maximally symmetric spaces with negative curvature i.e. Anti-de Sitter spaces.  We show that between  both scenarios there is a  {\it critical point} where the curvature blows up along the flow. Also the solutions for $d\geq 4$  satisfy the flow equations with Riemannian or pseudo-Riemannian metrics  due to the fact that the considered maximally symmetric spaces do not depend on time neither on the angular coordinates yielding equations that depend only on the radial coordinate and the Ricci flow parameter. Additionally we find an interesting effect of the flow consisting in a change of the signature of the metric when  passing the singular point. Besides the signature change, the sign of the curvature also experiences a transition from  positive to negative curvature, either with Riemannian or pseudo-Riemannian metrics throughout the singular point.     
\end{abstract}

\section{Introduction} 

The Ricci flow is a geometrical evolution equation that was originally proposed by R. S. Hamilton as  part of a program to classify $3$-manifolds \cite{Hamilton}. The Ricci flow has gained attention in physics since it behaves as a kind of heat equation for the metric. We can say, in a similar way, that as the heat equation homogenizes temperature in a wire, for instance, the Ricci flow homogenizes the metric on a manifold. If we  modify the Ricci flow by a family of diffeomorphisms we will get the so called Hamilton-DeTurck Ricci flow which is equivalent to the original formulation of the Ricci flow \cite{Turck}.
\footnote{For the reader who is interested in more formal  details of the work of Perelman and the Ricci flow we recommend to have a look at \cite{Perelman1}-\cite{morganprogres}.}

There are many  applications of the Ricci flow in the context of physics, as examples we can mention the following works \cite{Viqar}-\cite{Dai}. In \cite{Viqar} the Ricci flow was used to deform wormhole geometries within a numerical approach. As a result three scenarios for the evolution of the wormhole throat (shrinking, expanding or a steady throat) were found, which are determined by a critical parameter that identifies a family of wormhole geometries; this critical parameter also reveals a topological change on the manifold. In \cite{EnergyERF} an inequality  which relates the evolution of the area of a closed surface to its Hawking mass under the Ricci flow was derived, as a consequence, the rate of change of the area of a closed surface is bounded by its Hawking mass. The question of whether Perelman's entropy is related to the Bekenstein-Hawking geometric entropy from black hole thermodynamics was investigated in \cite{GFBHE} . By studying the fixed points of the flow, the authors concluded that Perelman's entropy is not connected to the geometric entropy, however, they proposed a modified flow which does appear to connect both entropies.

In \cite{wiseman} the Ricci flow was applied to four-dimensional Euclidean gravity with a $S^1\times S^2$ boundary, representing the canonical ensemble for gravity in a box. At high temperature the action has three saddle points: one of them is unstable under the Ricci flow. The other two lead to a large black hole and to a hot flat space, respectively, in the latter case via a topology-changing singularity. By using a maximum principle, in \cite{figueras} it was proved that Ricci solitons do not exist within static Lorentzian spacetimes which are asymptotically flat, Kaluza-Klein, locally AdS or have extremal horizons. 
In \cite{woolgar} the author reviewed how the Ricci flow is related to sigma models and Ricci solitons, and discussed the behaviour of the mass under the Ricci flow. Moreover, he constructed a Lorentzian (or Euclidean) $4d$ analytic Ricci soliton solution from a $3d$ seed soliton with the aid of a scalar field; this soliton represents a rare analytic configuration that can find applications of the Ricci flow in General Relativity. Later on, the behaviour of the asymptotically hyperbolic mass under the curvature-normalized Ricci flow of asymptotically hyperbolic and conformally compactifiable manifolds was investigated in \cite{BahelowskyWoolgar}: The mass turned out to exponentially decay to zero throughout the flow parameter for manifolds with $d\geq 3$.
In \cite{Dai} the change of the ADM mass of an asymptotic locally Euclidean (ALE) space along the Ricci flow was studied. The authors showed that the ALE property is preserved under the Ricci flow and  that the mass is invariant under the flow in three dimensions.

The Ricci flow also appears in physics as the renormalization group (RG) flow,  where the flow parameter describes the observation scale  of the considered theory.  It is common to refer to a  change in scale as a change in  the magnification of a microscope used to analyze such a theory \cite{Chowdhury}. This microscope will give us information about the physical  system, therefore the bigger the magnification  the more detailed will be the image. If we want to analyze the system with a higher magnification we will get  a lot of information. On the contrary,  if we use a small magnification we will get less information. In physics one can interpret the decrease of information as an increase of entropy, in some sense the Ricci flow decreases the information of a manifold and leaves just the  information needed to identify the manifold by its topology \cite{Chowdhury, Perelman1}.

Our motivation to study the  Ricci flow is to investigate the behaviour of de Sitter (dS) and Anti de Sitter (AdS) spaces under the Ricci flow. AdS and dS spaces  play a key role in physics, they are useful in cosmology and  other fields, for example in the AdS/CFT correspondence. One of our goals is to solve the modified Ricci flow equations for a simple ansatz of the DeTurck vector field.  We will see that under the Ricci flow, dS space shrinks  but  then, it surprisingly emerges as an AdS space. This unexpected behaviour is the result of an analytical extension of the metric for the dS spacetime along the Ricci flow.

Also our study is motivated by the fact that flat spaces are fixed points of the Ricci flow for some geometries. For example in \cite{Oliynyk} it was proved that if the Ricci flow evolves from specific initial conditions such as  rotationally symmetric, and asymptotically flat initial data, then the flow is immortal (i.e. the solution exists for the interval $(\lambda_0,\infty)$), remains asymptotically flat,  and converges to flat Euclidean space as the flow parameter diverges to infinity. Other examples where the flow converges to flat metrics for more complicated geometries,  are mentioned in \cite{RStability}, such as  direct-product metrics $(\mathcal{T}^2, \mu) \times (S^1, dx^2)$ with $\mu$ an arbitrary Riemannian metric on $\mathcal{T}^2$. 

We start our treatment of the Ricci flow in $3d$ deducing our solution from a spherically symmetric ansatz, this metric possesses just  one arbitrary function that must be determined and depends only on the flow parameter and the radial coordinate.  Our first solutions in $3d$ are a sphere and  hyperbolic space, these solutions can be generalized to $\mathbb{S}^d$ and $\mathbb{H}^d$. We also show that the  fixed points for these spaces are flat metrics. It is well  known that under the Ricci flow these spaces expand or shrink, a novel approach we adopt here is the addition of a timelike component into the metric which for obvious reasons is of interest in General Relativity.  This addition into the metric has three effects: the first one is that there are no more fixed points of the flow,  the second one is a change in the signature of the metric after passing the singular point, and the third one is a transition from positive to negative curvature throughout the singularity. A very interesting feature of our solution with a timelike component is that  the flow can be carried on either with Riemannian or pseudo-Riemannian metrics for all the $d\ge 4$ spaces.

\section{3d maximally symmetric solution to the Ricci flow}

The Ricci flow equation is given by
\begin{align}
\frac{\partial g_{\mu\nu}(\lambda)}{\partial 	\lambda} &=-2R_{\mu\nu}(g),   \hspace{0.5cm}  g(0) = g_0, \label{Riccif} 
\end{align}
here $R_{\mu\nu}(g)$ denotes the Ricci tensor depending on the metric $g_{\mu\nu}(\lambda)$, and $\lambda$ is the flow parameter. The main  idea behind the Ricci flow  is to find a mechanism that  allows the metric tensor $g$ evolve under a partial differential equation which resembles the heat equation. Thus, as the $\lambda$ parameter goes on, the metric tensor becomes homogeneous throughout the flow.

In some cases it is useful to work with a modified version of the Ricci flow given by the Hamilton-DeTurck Ricci flow. D. DeTurck used the known "DeTurck trick"  which makes use of a family of diffeomorphisms to modify the Ricci flow  \cite{Turck}. The Hamilton-DeTurck Ricci flow is defined by the equation
\begin{equation}
\partial_\lambda g_{\mu \nu} =-2R_{\mu \nu} +  \nabla _\mu V_\nu+ \nabla _\nu V_\mu, 
\label{ric}
\end{equation}
where $g_{\mu \nu}$ identifies a Riemannian metric, $\lambda$ is the flow parameter, $V_\mu$  is the DeTurck vector field which generates diffeomorphisms along the flow, and $R_{\mu \nu}$ is the Ricci  tensor. 

We  are going to solve the Hamilton-DeTurck Ricci  flow in 3$d$ starting with a spherically symmetric  ansatz, so the  DeTurck vector field will be also spherically symmetric. For the metric we will use
\begin{equation}
ds ^2= d \rho^2 + P(\rho, \lambda) d\Omega ^2, \label{metric}
\end{equation}
and for the  DeTurck vector field
\begin{equation}
V_{\rho}=w_1^2(\lambda)\rho. 
\label{diff}
\end{equation}

Then the Hamilton-DeTurck Ricci flow equations (\ref{ric}) render just two relevant equations\footnote{The $\theta\theta$ and $\phi\phi$ components yield the same differential equation.}
\begin{eqnarray}
0&=&-2\frac{P'^2-2 P P''}{2 P^2} + 2V'_{\rho},\label{rhorho}\\
\dot{P}&=&-2\left(1-\frac{P''}{2}\right)+P'V_{\rho}, \label{thetatheta}
\end{eqnarray} 
that can be rewritten as 
\begin{align}
\dot{P}=P''+V_{\rho}P'-2, \label{diynamicp}\\
V_{\rho}'=-\frac{P''}{P}+\frac{P'^2}{2P^2}, 
\label{constriction}
\end{align}
where a dot denotes derivatives with respect to the Ricci flow parameter $\lambda$ and primes stand for derivatives with respect to the radial coordinate $\rho$.
The first equation is a dynamical equation for $P$, while the second one is a constriction equation that appears due to the fact that the $g_{\rho\rho}$ metric component of (\ref{metric}) does not depend on the flow parameter $\lambda$. These equations are valid for any spherically symmetric DeTurck vector field.

Solving for $P$ in (\ref{constriction}) with DeTurck vector field ansatz (\ref{diff}) we obtain 
\begin{equation}
P=w_3(\lambda) \cos ^2\left[\frac{w_1(\lambda)}{\sqrt{2}}\left(\rho-2w_2(\lambda) \right) \right]. \label{solP}
\end{equation}
Here $w_1(\lambda), w_2(\lambda), w_3(\lambda)$ are functions of $\lambda$ that we are going to determine using equation (\ref{diynamicp}). By substituting  solution (\ref{solP}) into  equation (\ref{diynamicp}) and making use of some  trigonometric identities it is not difficult to see that the $w_i$ functions must be 
\begin{equation}
w_1=\frac{1}{\sqrt{2(\lambda_0-\lambda)}}, \hspace{1.5cm} w_2=k\sqrt{2(\lambda_0-\lambda)}, \hspace{1.5cm} w_3=4(\lambda_0-\lambda),
\end{equation}
with $k$ and $\lambda_0$ arbitrary constants.  Thus the solution can be written as
\begin{equation}
P(\lambda,  \rho )=4(\lambda_0-\lambda)\cos ^2\left[ \frac{\rho}{2\sqrt{\lambda_0-\lambda}}+k_0  \right], \label{solutionP}
\end{equation}
with $k_0=-\sqrt{2}\,k$ an arbitrary constant.  The fixed points are given by the equation 
$$ 
\frac{\partial g(\lambda)}{\partial \lambda}=0\, ,
$$ 
this implies  
$$
\cos\left(\frac{\rho}{2\,\sqrt{\lambda_0-\lambda}}+k_0\right) +\frac{\rho}{2\,\sqrt{\lambda_0-\lambda}} \sin\left(\frac{\rho}{2\,\sqrt{\lambda_0-\lambda}}+k_0\right)=0. \label{fijos}
$$
To satisfy this equation for every $\rho$ we require $k_0=\frac{(2n+1)\pi}{2}$, with $n$ an integer number, and take the limit $\lambda \rightarrow - \infty$. By making this election for $k_0$, our solution turns into
\begin{equation}
P=4(\lambda_0-\lambda)\sin^2 \left[\frac{\rho}{2\sqrt{\lambda_0-\lambda}} \right]. \label{Psin}
\end{equation}
It can be seen that when $\lambda \rightarrow  -\infty$ the metric (\ref{metric}) reduces to $ds^2= d\rho^2+\rho^2 d\Omega ^2$;  thus, the fixed point along the Ricci flow for negative $\lambda$ is an asymptotic flat spacetime. 

So far we have considered  that $\lambda$ is in the interval $(-\infty,\lambda_0)$, now consider $\lambda \in (\lambda_0, \infty)$, in this case the argument of the function $\sin(\cdot)$ in equation (\ref{Psin}) becomes imaginary, then we can use the identity
$ \text{sin}(\mathtt{i} z) =\mathtt{i} \text{sinh} (z), \Rightarrow  \text{sin}^2(\mathtt{i} z) =-\text{sinh}^2(z)$ to analytically expand the metric (\ref{metric}) with the solution (\ref{Psin}) along the Ricci flow; thus, the obtained line element can be written as
\begin{equation}
ds^2=d \rho ^2+4(\lambda-\lambda_0)\sinh^2 \left[  \frac{\rho}{2\sqrt{\lambda-\lambda_0}} \right]d\Omega ^2,
\end{equation}
which also reduces to $ds^2= d\rho^2+\rho^2 d\Omega ^2$ when $\lambda \rightarrow  +\infty$, indicating one more time that the fixed point for positive $\lambda$ along the Ricci flow corresponds to an asymptotic flat metric as well.

It is also interesting to note that our solutions are maximally symmetric spaces. The equation for a $d$ dimensional maximally symmetric space, at any point, in any coordinate system reads 
\begin{equation}
R_{\rho \mu \sigma \nu} = \frac{R}{d(d-1)}\left( g_{\rho \sigma}g_{ \mu \nu}-g_{\rho \nu}g_{ \mu \sigma }\right). \label{maximally}
\end{equation}
In our $3d$ case the Ricci curvature is given by
\begin{equation}
R(\lambda)=\frac{3}{2(\lambda_0-\lambda)},
\label{R}
\end{equation} 
therefore, the curvature tends to zero for  $\lambda \rightarrow \pm \infty$, revealing the asymptotic fixed points of the metric along the Ricci flow.

Thus, the curvature can be positive or negative depending  on our election of $\lambda$ and $\lambda_0$. The curvature is positive when $\lambda$ is in the range $\left(-\infty, \lambda_0 \right)$  and negative in the range $\left(\lambda_0, \infty \right)$.  At the critical point $\lambda=\lambda_0$, the curvature blows up. 
These solutions are spheres and  hyperbolic spaces and can be generalized to $\mathbb{S}^d$ and $\mathbb{H}^d$, the extension to $d$ dimensions will not affect the structure of the solution  and  the only change  will be reflected in $d\Omega^2_d$. One can see that  these solutions are in fact  spheres and hyperbolic spaces multiplied  by a  factor proportional to $\lambda_0-\lambda$ using the coordinate transformation $r=\frac{\rho}{2\sqrt{\lambda_0-\lambda}}$.

\section{Solution in 4 dimensions}

At this point we know the key features of  the geometrical properties of the solution in 3$d$. It will be of interest  to ask what happens if we want to extend the solution to a $4d$ metric with a  timelike component. Let us remember that an universal covering of AdS space is given by \cite{thermalgauge}
\begin{equation}
ds^2=-l^2\cosh^2 \left(\frac{v}{l} \right)dt^2+dv^2 +l^2 \sinh^2\left(\frac{v}{l} \right)d\Omega_{d-2}, \label{covering}
\end{equation}
where $d$ is the  space-time dimension and $l$ is a parameter. In fact the metric (\ref{covering}) is a solution to the Hamilton's DeTurck Ricci flow. 
Our solution in $3d$ has exactly the same form as the spatial part of (\ref{covering}) if we identify $v$ with $\rho$ and $l^2$ with $4(\lambda_0-\lambda)$. Now we are going to  find a solution in 4$d$ performing an analogue treatment to the 3$d$ case. Let us start with the ansatz
\begin{equation}
ds^2= Q(\lambda, \rho)dt^2 + d\rho^2 + P(\lambda, \rho)d\Omega ^2,  \hspace{0.5cm} \text{and} \hspace{0.5cm} V_{\rho}(\lambda, \rho)=w_1^2(\lambda) \rho + u(\lambda),
\end{equation}
then the flow equations can be written as
\begin{align}
\dot{Q} &=\frac{P'Q'}{P}-\frac{Q'^2}{2Q}+Q'' +Q'\left( w_1^2\rho+u \right), \label{dotq}\\
0 &=-\frac{P'^2}{P^2}+\frac{2P''}{P}-\frac{Q'^2-2QQ''}{2Q^2}+2w_1^2, \label{constrictionPQ}\\
\dot{P} &=-2+\frac{P'Q'}{2Q}+P''+P'\left( w_1^2 \rho+u \right). \label{dotp}
\end{align}

We can solve equation (\ref{constrictionPQ}) by setting
\begin{align}
\frac{P'^2}{P^2}-\frac{2P''}{P}&=\frac{4}{a}w_1^2,\label{p}\\
\frac{Q'^2}{Q^2}-\frac{2Q''}{Q}&=\frac{4(a-2)}{a}w_1^2 \label{q}, 
\end{align}
where $a$ is a constant different from $2$ and $0$ . The solutions  for $P$ and $Q$ in equations (\ref{p}) and (\ref{q}) are 
 \begin{align}
 P&=\text{cos}^2\left[ \frac{w_1}{\sqrt{a}}\rho - w_2 \right]w_3, \label{pcos}\\
 Q&=\text{cos}^2\left[ \sqrt{\frac{(a-2)}{a}}w_1\rho - w_4 \right]w_5;\label{qcos}
 \end{align}
by substituting (\ref{pcos}) and (\ref{qcos}) into  (\ref{dotp}) we obtain the following restriction
\begin{equation}
\begin{split}
&\dot{w}_3 \, \text{cos}^2x -2w_3 \left[\frac{\dot{w}_1}{\sqrt{a}}\rho   - \dot{w}_2 \right] \text{cos}x \,\text{sin}x  = -2+2\frac{\sqrt{a-2}w_1^2w_3}{a}\text{tan}y\, \text{cos}x\, \text{sin}x \,+\\ &2\frac{w_1^2w_3}{a}\left[\text{sin}^2x - \text{cos}^2x  \right] 
-2\frac{w_1w_3}{\sqrt{a}}\left[ w_1^2\rho + u  \right]\text{cos}x\,\text{sin}x,
\end{split} \label{tany}
\end{equation}
where

\begin{equation}
x=\frac{w_1}{\sqrt{a}}\rho - w_2,  \hspace{0.5cm} y= \sqrt{\frac{a-2}{a}}w_1\rho - w_4.
\end{equation}

In order to satisfy (\ref{tany}) we  need to gather  similar terms and then cancel  the coefficients  that multiply  the  trigonometric functions, nonetheless  in the first line of (\ref{tany}) the argument of the tangent function  is not the same of most of the other trigonometric functions. We can overcome this difficulty by setting
\begin{equation}
y=x-\frac{(2n+1)}{2}\pi,
\end{equation}
 with $n$ an integer, this condition  determines the value of $a=3$ and  fixes $w_4=w_2+\frac{(2n+1)}{2}\pi$. This relation  also implies that  
\begin{equation}
\text{tan}y=-\text{cot}x, \label{tancot}
\end{equation}
 hence we can rewrite equation (\ref{tany}) as
\begin{equation}
\begin{split}
&\left[\dot{w}_3 +  2 w_1^2w_3  \right]\text{cos}^2 x + 2\frac{w_3 \rho}{\sqrt{3}}\left[-\dot{w}_1+w_1^3   \right]\text{cos}x\,\text{sin}x \,+\\
&2w_3 \left[ \dot{w}_2 +\frac{w_1 u}{\sqrt{3}} \right] \text{cos}x\,\text{sin}x+\left[ 2-\frac{2w_1^2w_3}{3}  \right]=0.
\end{split}
\end{equation}
This condition yields the following equations
\begin{equation}
\dot{w}_3 +  2 w_1^2w_3 =0, \hspace{1.5cm}
\dot{w}_1-w_1^3 =0, \hspace{1.5cm}
\dot{w}_2 +\frac{w_1 u}{\sqrt{3}}=0, \hspace{1.5cm}  1-\frac{w_1^2w_3}{3} =0.\label{w1}
\end{equation}

Similarly, by substituting (\ref{pcos}) and (\ref{qcos}) into  (\ref{dotq}) and using (\ref{tancot}) we obtain another restriction
\begin{equation}
\begin{split}
\frac{2w_5\rho}{\sqrt{3}}\left[w_1^3-\dot{w}_1\right]\text{cos}y\,\text{sin}y +2w_5\left[\dot{w}_4 +\frac{w_1u}{\sqrt{3}}  \right]\text{cos}y\,\text{sin}y\, + \left[\dot{w}_5 +2w_1^2w_5   \right]\text{cos}^2y\,=0,
\end{split}
\end{equation}
from this equation it follows that
\begin{align}
\dot{w}_1-w_1^3 &=0, \hspace{1.5cm}
\dot{w}_4 +\frac{w_1u}{\sqrt{3}}=0, \hspace{1.5cm} \dot{w}_5 +2w_1^2w_5=0,\label{w2}
\end{align}
by solving the equations (\ref{w1}) and (\ref{w2}) for the $w_i$ functions we obtain
\begin{align}
w_1=\frac{1}{\sqrt{2(\lambda_0-\lambda)}},\hspace{1.5cm}
w_3=6(\lambda_0-\lambda),\hspace{1.5cm}
w_5=\kappa(\lambda_0-\lambda),
\end{align}
where $\kappa$ arises as an arbitrary integration constant from the last condition of (\ref{w2}) once $w_1$ has been obtained. We have just one equation for $w_2$ and $w_4$  due to the fact  that $w_4 = w_2 + \frac{(2n+1)}{2}\pi$, this equation is
\begin{equation}
\dot{w_2}+\frac{u}{\sqrt{6(\lambda_0-\lambda)}}=0,
\end{equation}
where $u$ is an arbitrary function of $\lambda$, which yields in turn $w_2$ as an arbitrary function of $\lambda$ that we will denote henceforth as $\tilde{w}$. Now we can write the line element  in $4$ dimensions for $\lambda<\lambda_0$
\begin{equation}
ds^2=\kappa(\lambda_0-\lambda)\text{cos}^2\left[ \frac{\rho}{\sqrt{6(\lambda_0-\lambda)}} -\tilde{w} \right]dt^2+ d\rho ^2
+ 6(\lambda_0-\lambda)\text{sin}^2\left[ \frac{\rho}{\sqrt{6(\lambda_0-\lambda)}} -\tilde{w}   \right]d\Omega^2, \label{general4d}
\end{equation}
with $\kappa$ an arbitrary constant and $\tilde{w}(\lambda)$  an arbitrary function of $\lambda$.  

Note that the solution (\ref{general4d}) is well defined for $\lambda \in (-\infty, \lambda_0)$, of course we can analytically extend this solution to the  interval $\lambda \in (\lambda_0,\infty)$ as in the $3d$ case.

For the $\lambda >\lambda_0$ case we have
\begin{equation}
ds^2=-\kappa(\lambda-\lambda_0)\text{cosh}^2\left[ \frac{\rho}{\sqrt{6(\lambda-\lambda_0)}} -\tilde{w} \right]dt^2+ d\rho ^2
+ 6(\lambda-\lambda_0)\text{sinh}^2\left[ \frac{\rho}{\sqrt{6(\lambda-\lambda_0)}} -\tilde{w}   \right]d\Omega^2. \label{general4dh}
\end{equation} 

Note as well that the $\kappa$ constant is completely arbitrary, thus $\kappa$ can be positive or negative, which implies that  the flow will exist either with Riemannian or pseudo-Riemannian metrics for this concrete static class of solutions.  

Although the nature of the equations corresponding to the Ricci flow in general is not the same for  Riemannian or pseudo-Riemannian metrics, since they drastically change from a heat-like equation to a hyperbolic one, for the specific case of maximally symmetric spaces we have found a solution that works for both signatures by construction. This is due to the fact that the considered maximally symmetric spaces do not depend on the time and angular coordinates, leading to equations that depend only on the radial coordinate together with the Ricci flow parameter $\lambda$. 

Also note that, if we start the  flow with a determined signature, we can see that due to the structure of the solution, after passing the singularity the metric will suffer a  {\it signature change}. To be more specific when  $\lambda$ exceeds $\lambda_0$, $\lambda_0 -\lambda$ becomes negative and $\text{sin}^2(\mathtt{i}z)$ in (\ref{general4d}) becomes $-\text{sinh}^2(z)$. Thus the change in the sign of $\lambda_0-\lambda$ is absorbed by the minus sign of  $-\text{sinh}^2(\cdot)$, so that the angular metric components preserve the sign when $\lambda_0-\lambda$ becomes negative. Conversely, in the case  of the $g_{tt}$ component, the squared cosine function  does not change  its sign whereas the factor $\lambda_0-\lambda$ does when $\lambda$ becomes greater than $\lambda_0$. Having this in mind  and the fact that the flow can be carried on with a Riemannian or pseudo-Riemannian metric we can say that the flow  will always exhibit a "transition" from a Riemannian to a pseudo-Riemannian metric or vice versa, no matter which one is deformed first, again for the concrete family of static solutions considered in this work.

\begin{figure}
\centering
\includegraphics[width=5cm]{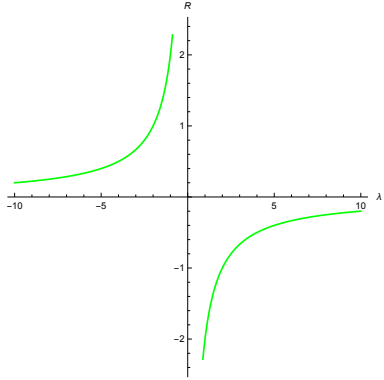}
\caption{Plot of the scalar curvature $R$ as a function of $\lambda$.   Along the evolution of the Ricci flow the curvature invariant blows up when $\lambda$ reaches $\lambda_0$. Here we have chosen $\lambda_0=0$ for simplicity. } 
\label{riccicurvature}
\end{figure}

The curvature in $4d$ is given by
\begin{equation}
R=\frac{2}{\lambda_0-\lambda},
\end{equation}
thus, it is evident that this invariant blows up when $\lambda=\lambda_0$, i.e. when the transition takes place along the Ricci flow, see Fig. \ref{riccicurvature}. 

It is worth mentioning here that although we have a solution for the flow, now there is no choice of $\lambda$ and $\tilde{w}$ that satisfies the system of equations  
$\frac{\partial g_{\mu\nu}}{\partial\lambda}=0$. We see that unlike the $3d$ (spacelike) case, where the system $\frac{\partial g_{\mu\nu}}{\partial\lambda}=0 $ contains just one relevant equation, in the $4d$ case we have two equations that must be simultaneously satisfied:
\begin{equation}
\frac{\partial Q}{\partial \lambda}=0 \label{q},
\end{equation}
\begin{equation}
\frac{\partial P}{\partial \lambda}=0\label{pl};
\end{equation}
by substituting $P$ and $Q$ into these equations we obtain
\begin{equation}
\cos\left(\frac{\rho}{\sqrt{6(\lambda_0-\lambda)}} - \tilde{w}\right) + \left[ \frac{\rho}{\sqrt{6(\lambda_0-\lambda)}} - 2(\lambda_0-\lambda)\dot{\tilde{w}} \right]
\sin\left(\frac{\rho}{\sqrt{6(\lambda_0-\lambda)}} - \tilde{w}\right)=0 
\label{Q=0},
\end{equation}
\begin{equation}
-\sin\left(\frac{\rho}{\sqrt{6(\lambda_0-\lambda)}} - \tilde{w}\right) + \left[ \frac{\rho}{\sqrt{6(\lambda_0-\lambda)}} - 2(\lambda_0-\lambda)\dot{\tilde{w}} \right]
\cos\left(\frac{\rho}{\sqrt{6(\lambda_0-\lambda)}} - \tilde{w}\right)=0 
\label{P=0};
\end{equation}
which in turn render the following relations

 
\begin{equation}
\frac{\rho}{\sqrt{6(\lambda_0-\lambda)}}-2(\lambda_0-\lambda)\dot{\tilde{w}}=-\cot\left(\frac{\rho}{\sqrt{6(\lambda_0-\lambda)}} - \tilde{w}\right), 
\label{cot}
\end{equation}
\begin{equation}
\frac{\rho}{\sqrt{6(\lambda_0-\lambda)}}-2(\lambda_0-\lambda)\dot{\tilde{w}}=\tan\left(\frac{\rho}{\sqrt{6(\lambda_0-\lambda)}} - \tilde{w}\right).
\label{tan}
\end{equation}
These equations can not be satisfied simultaneously by $\tilde{w}$ for an arbitrary $\rho$ since their RHS never coincide, therefore we conclude that for this case there are no fixed points.

\section{Solution in $d > 4$ dimensions}

It is interesting to note that  the $4d$ solution  can be generalized to the $d > 4$ case. If we add more dimensions the only change will be a modification in the dimension of the angular sector of the metric $d\Omega_{d-2}$ and we will have  similar solutions for $P$ and $Q$. As we have shown, the line element of a maximally symmetric space  can be written as $(\ref{covering})$, therefore that  will be a solution for the Ricci flow in $d$ dimensions with $\rho=v$ and $l^2=2(d-1)(\lambda_0-\lambda)$.

The flow equations in $d$ dimensions are given by
\begin{equation}
 \begin{split} 
 \dot{P}=& -2(d-3)+\frac{P'Q'}{2Q}+P''+\frac{(d-4)}{2}\frac{P'^2}{P}   +\left(w_1 ^2\rho+u \right)P',\\
 0=&-(d-2)\frac{P'^2}{P^2}+2(d-2)\frac{P''}{P}-\frac{Q'^2-2QQ''}{Q^2}+2w_1^2,\\
 \dot{Q}=&\frac{(d-2)}{2}\frac{P'Q'}{P}-\frac{Q'^2}{2Q}+Q''+(w_1^2\rho+u) Q'.
 \end{split}
 \end{equation}
 with $V_{\rho}(\lambda,\rho)=w_1 ^2(\lambda)\rho+u(\lambda)$, where $u(\lambda)$ is an arbitrary function of $\lambda$. Following an analogue process to the case $d=4$, it is possible to see that  a solution for this system is given by
 \begin{equation}
 \begin{split}
Q&=\kappa(\lambda_0-\lambda)\text{cos}^2\left[ \frac{\rho}{\sqrt{2(d-1)(\lambda_0-\lambda)}}-\tilde{w}  \right], \\
 P&=2(d-1)(\lambda_0-\lambda)\text{sin}^2\left[ \frac{\rho}{\sqrt{2(d-1)(\lambda_0-\lambda)}}-\tilde{w}  \right] ,\\
  w_1^2&=\frac{1}{2(\lambda_0-\lambda)},
\end{split} 
 \end{equation}
 with $\kappa$ an arbitrary constant, $\tilde{w}$ an arbitrary function of $\lambda$ and $\lambda<\lambda_0$.
 
This solution can be analytically extended  for the  case $\lambda>\lambda_0$. In that case the functions $Q$ and $P$ are
\begin{equation}
\begin{split}
Q&=\kappa(\lambda-\lambda_0)\text{cosh}^2\left[ \frac{\rho}{\sqrt{2(d-1)(\lambda-\lambda_0)}}-\tilde{w}  \right], \\
 P&=2(d-1)(\lambda-\lambda_0)\text{sinh}^2\left[ \frac{\rho}{\sqrt{2(d-1)(\lambda-\lambda_0)}}-\tilde{w}  \right]. \\
\end{split} 
\end{equation} 
Then in $d$ dimensions the line element can be written  as
 \begin{equation}
ds^2= \kappa(\lambda_0-\lambda)\text{cos}^2 \left(\frac{\rho}{l} \right)dt^2+d\rho^2+l^2\text{sin}^2\left( \frac{\rho}{l}  \right)d\Omega_{d-2}^2,\label{sch1}
\end{equation}
for $\lambda<\lambda_0$ and 
 \begin{equation}
ds^2= -\kappa(\lambda-\lambda_0)\text{cosh}^2 \left(\frac{\rho}{l} \right)dt^2+d\rho^2+l^2\text{sinh}^2\left( \frac{\rho}{l}  \right)d\Omega_{d-2}^2,\label{sch2}
\end{equation} 
for $\lambda_0<\lambda$, with $l^2=2(d-1)(\lambda-\lambda_0)$.
Here $\kappa$ plays the same role as in the $4d$ family of metrics. It can also be shown that there are no fixed points  for $d > 4$ since there is no choice of $\tilde{w}$ or $\lambda$ that satisfies the conditions $\frac{\partial g_{\mu\nu}}{\partial \lambda}=0$.

From these  metrics it can be noted  that  the change in the signature metric of maximally symmetric spaces across a critical point is a universal feature of such spaces;  in this transition Riemannian metrics are transformed into pseudo-Riemannian metrics  and vice versa. This can be seen either from (\ref{sch1}) or (\ref{sch2}) thanks to the factor $\lambda_0-\lambda$ in the $g_{tt}$ components of these metrics.

These metrics will look like flat spacetime for $|\lambda|$ sufficiently large. Suppose we choose a slice of the flow by fixing a $\lambda_*$ sufficiently large such that $\frac{1}{\lambda_0-\lambda_*} \rightarrow 0$, and a $\tilde{w}(\lambda)$ that decays faster than $\frac{1}{\lambda_0-\lambda_*} $,  then the metric (\ref{sch1}) reads

\begin{equation}
\begin{split}
&ds^2=\kappa(\lambda_0-\lambda_*)\cos^ 2\left(\frac{\rho}{\sqrt{2(d-1)(\lambda_0-\lambda_*)}}-\tilde{w} \right)dt^2+ d\rho^2 +\\
& 2(d-1)(\lambda_0-\lambda_*)\sin^2\left(\frac{\rho}{\sqrt{2(d-1)(\lambda_0-\lambda_*)}} -\tilde{w}\right)d\Omega_{d-2}^2 
\end{split}
\end{equation}
and becomes
\begin{equation}
ds^2 \approx \kappa(\lambda_0-\lambda_*)dt^2 + d\rho^2 +\rho^2d\Omega^2_{d-2}\,;
\end{equation}
this is a flat metric since the constant $\kappa(\lambda_0-\lambda_*)$ can be absorbed into the time coordinate. For successive slices with $|\lambda| >|\lambda_*|$  the local geometry remains flat.
Thus, we can say that regardless of whether there are fixed points,  locally  the  spacetime will  look flat for sufficiently large $\lambda$. A similar analysis can be performed in the language of the scalar curvature depending on the flow parameter $\lambda$.

\section{The solution by choosing a different metric ansatz}

In this section we are going to reproduce the solutions obtained in equation (\ref{sch1})-(\ref{sch2}) by using a convenient choice of the coordinates and the Ricci flow instead of  the modified DeTurck flow. 
Let $\hat{g}_{\mu\nu}$ be a maximally symmetric space, thus it satisfies equation (\ref{maximally}), here we use the symbol $\,\hat{}\,$ for quantities that do not depend on $\lambda$. By contracting the first and third indices of the Riemann tensor in (\ref{maximally}) we obtain $\hat{R}_{\mu\nu}=\frac{\hat{R}}{d}\hat{g}_{\mu\nu}$. Also for a maximally symmetric space it can be seen that $\hat{R}=\alpha\,d(d-1)$, where $\alpha$ is the inverse of the radius of curvature,  thus we have for the Ricci tensor
\begin{equation}
\hat{R}_{\mu\nu}=\alpha (d-1) \hat{g}_{\mu\nu}.
\end{equation}
Now we are going to write an ansatz for the  metric that will be deformed under the Ricci flow 
\begin{equation}
g_{\mu\nu}=F(\lambda)\hat{g}_{\mu\nu}, \label{scaling}
\end{equation}
from this equation one can see that for this particular case $R_{\mu\nu}=\hat{R}_{\mu\nu}$. Substituting this into the  Ricci flow  $\partial_{\lambda} g_{\mu \nu}=-2R_{\mu\nu}$,  we have
\begin{equation}
\begin{split}
\partial_{\lambda} \left(F(\lambda)\hat{g}_{\mu \nu}\right)&=-2\alpha\,(d-1)\hat{g}_{\mu\nu},\\
\end{split}
\end{equation}
then the solution for $F(\lambda)$ is
\begin{equation}
F(\lambda)= 2\alpha(d-1)(\lambda_0-\lambda).
\end{equation}
This result is independent  of the metric signature.
Now we  make a choice for the coordinate system of the maximally symmetric space $\hat{g}_{\mu\nu}$ 
\begin{equation}
 \hat{g}_{\mu \nu}= \hat{Q}(\hat{\rho})dt^2+ d\hat{\rho}^2+\hat{P}(\hat{\rho})d\Omega^2, \label{pq}
 \end{equation} 
and by considering equation (\ref{scaling}) and (\ref{pq}) we have
\begin{equation}
Q(\rho)dt^2+ d\rho^2+P(\rho)d\Omega^2=F(\lambda)\left(\hat{Q}(\hat{\rho})dt^2+ d\hat{\rho}^2+\hat{P}(\hat{\rho})d\Omega^2\right),
\end{equation}
consequently
\begin{equation}
\frac{\partial \rho}{\partial \hat{\rho}}=\sqrt{F(\lambda)},
\end{equation}
whose solution is
\begin{equation}
\rho=\sqrt{F(\lambda)}\left(\hat{\rho}+G(\lambda)\right),
\end{equation}
where $G(\lambda)$ is an arbitrary function of $\lambda$.
By substituting $\hat{\rho}$ into  $\hat{Q}(\hat{\rho})$ we have
\begin{equation}
\begin{split}
Q(\rho)=F(\lambda)\hat{Q}\left(\frac{\rho}{\sqrt{F(\lambda)}}-G(\lambda) \right).
\end{split} \label{gsolution}
\end{equation}
A similar expression holds for the pair of functions $P(\rho)$ and $\hat{P}(\hat{\rho})$. 
Since this coordinate transformation depends on $\lambda$, the Ricci flow equation is transformed into the modified DeTurck Ricci flow.

This equation reparameterizes the metrics that we have obtained in previous sections; it includes  all maximally symmetric spaces deformed by the DeTurck flow. The solutions found earlier are particular cases in $3$, $4$ and $d$ dimensions  with appropriate $P(\rho)$ and $Q(\rho)$ for the choice of the  diffeomorphism (\ref{diff}).

\section{Discussion}

In this work  we have found explicit solutions to the  Ricci flow for  an arbitrary dimension, particularizing for the  $3d$ and $4d$ cases. These solutions are  maximally symmetric spaces and  therefore are solutions to the Einstein equations with a cosmological constant (either positive or negative). We have found that the Ricci curvature is constant along the Ricci flow in the sense that by fixing the value for the flow parameter, we get a defined constant value for the curvature.  The flow develops a singularity at the finite value $\lambda=\lambda_0$ of the flow parameter. This  point divides  spaces with positive  constant curvature (for $\lambda_0-\lambda>0$) from those with negative constant one ( for $\lambda-\lambda_0>0$) connected via an analytical extension of the metric.

In this sense, the metric solution experiences a "transition"  at the critical point.  The behaviour of the flow can be summarized as follows: by starting from a maximally symmetric space with positive curvature, the flow increases the curvature  at the same time as the  space is contracting until it shrinks to a point. This point is the critical value where the curvature blows up. Then the space  emerges as  a maximally symmetric space, this time with a large negative curvature, and evolves towards a zero curvature space. In the case of the $3d$ space, we see that it indefinitely expands until reaching a fixed point of the flow. We have shown that the fixed points of the flow in $3d$ are spaces that locally can be represented  with a flat metric. 
The $3d$ cases are the well known sphere and hyperbolic spaces evolving under the Ricci flow and can be generalized to $d$ dimensions. 

The $4d$ solution also can be generalized to arbitrary $d$ dimensions but is different from the $3d$ case and its generalizations because we have added a timelike component to the metric. As a consequence we have three effects,  one of them is that there are no fixed points for the flow, the second  is  a change in the signature of the metric when passing the critical value, and the third one is a transition from positive to negative curvature throughout the singular point. As we have shown, passing trough the critical point from positive to negative curvature yields  a change in the signature of the metric. In other words, it was shown that  we can start with a Riemannian metric  and  by means of the flow,  after passing a singularity,  we  will obtain a pseudo-Riemannian metric.  On the other hand, by starting with a pseudo-Riemannian metric we will obtain a Riemannian metric after reaching the singular point. 

This reasoning is supported by the idea that the solution to the flow is one connecting two spaces. To support this claim it would be very interesting to find a physical system which could be modeled by this "transition" as the phase transitions that take place within the framework of the AdS/Condensed Matter Physics correspondence \cite{AdSCMT},   in which a state of the system evolves under the change of an external parameter, say temperature, doping or pressure; when the external parameter reaches a critical value, described by the singular point of the curvature, the condensed matter system drastically changes its phase, transforming into a completely different state of matter; however for the time being we can state this reasoning just as a conjecture. 

We would like to close this Section by addressing the issue of the existence of Ricci flow fixed points for the $d\ge 4$ spacetimes considered in this work. By taking into account that there is only a small class of genuine fixed points of the Ricci flow, here we have attached ourselves to the classical definition that requires the metric to obey the relation $\frac{\partial g_{\mu\nu}}{\partial\lambda}=0$. 

However, there are in the literature some generalized definitions of a fixed point. For instance, a Riemannian manifold is a fixed point of the (unnormalized) Ricci flow (\ref{Riccif}) if and only if the metric $g_{\mu\nu}$ is Ricci flat, whereas a compact manifold is a fixed point of the normalized Ricci flow if and only if $g_{\mu\nu}$ is an Einstein metric \cite{ChowKnopf}. Alternatively, the useful concept of quasi-convergence was also coined in \cite{HamiltonQ} and rigorously defined in \cite{Knopf} by establishing an equivalence relation between two evolving Riemannian metrics on a given manifold. 
However, although the  $\frac{1}{\lambda}$  decay in the curvature seems to be a common behaviour for some homogeneous metrics \cite{HamiltonQ}, there is no definition that compares this behaviour  for different manifolds.
In the definition of quasi-convergence given in \cite{ChowKnopf} the aforementioned metrics are defined and compared in the same manifold. Since Minkowski  and  de Sitter (or Anti-de Sitter) spaces represent different manifolds, we can not compare these metrics according to this definition of quasi-convergence. 


We finally would like to point out that it would be interesting to explore the possibility of giving a less restrictive definition of a fixed point of the Ricci flow that would encompass the situations that we have found in the present paper, inasmuch as our maximally symmetric spacetimes tend to flat metrics as $\lambda \rightarrow \pm \infty$, even if in these limits the classical definition of a fixed point breaks down.

\section{Acnowledgments}
The  work  of  A.H.A.  was  completed  at  the  Aspen  Center for Physics, which is supported by National Science Foundation grant PHY-1607611 and a Simons Foundation grant as well. He expresses his gratitude to the ACP for providing an inspiring and encouraging atmosphere for conducting part of this research. R.C.F. and A.H.A. acknowledge a VIEP-BUAP grant and thank SNI for support. J.A.O.S thank CONACYT for support. The authors thank U. Nucamendi and T. Oliynyk for valuable discussions and useful comments. The authors would like to thank as well an anonymous referee for wise suggestions on improving the paper.

\bibliographystyle{unsrt}

\end{document}